\documentclass[journal]{IEEEtran}
\usepackage{hyperref}
\usepackage{tablefootnote}
\usepackage{url}
\usepackage{flushend}
\usepackage{enumitem}	
\usepackage{amssymb}
\usepackage{xcolor}
\usepackage{pgf,interval}
\usepackage{graphicx} 
\usepackage{ifthen}
\usepackage{multirow}
\usepackage{booktabs}
\usepackage{tikz}
\usepackage{csquotes}
\usepackage{mathtools}
\usepackage[flushleft]{threeparttable}
\usepackage{multirow}

\newboolean{showcomments}
\setboolean{showcomments}{false}
\ifthenelse{\boolean{showcomments}}
{ \newcommand{\mynote}[3]{
     \fbox{\bfseries\sffamily\scriptsize#1}
        {\small$\blacktriangleright$\textsf{\emph{\color{#3}{#2}}}$\blacktriangleleft$}}}
{ \newcommand{\mynote}[3]{}}
\definecolor{asparagus}{rgb}{0.53, 0.66, 0.42}

\newcommand{\fh}[1]{\mynote{FH}{#1}{red}}

\usepackage{balance}
\usepackage[binary-units=true]{siunitx} 
\usepackage[caption=true]{subfig}
\ifCLASSOPTIONcompsoc
  \usepackage[nocompress]{cite}
\else
  \usepackage{cite}
\fi

\ifCLASSINFOpdf
\else
\fi

\ifCLASSINFOpdf
\else
\fi
\begin{document}
%
\title{Machine-Learning-Assisted Hybrid Cooling of a Raspberry Pi 5}
%
%
%

\author{Abdallah~Ghazal, Fazal~Hameed, Elshafey~Elshafey, Mohammad~Awadallah, and Alhasan~Emam%
\thanks{A. Ghazal, E. Elshafey and A. Emam are with the Department of Mechanical Engineering, American University of Sharjah, Sharjah, United Arab Emirates.}%
\thanks{F. Hameed, and M. Awadallah are with the Department of Computer Science and Engineering, American University of Sharjah, Sharjah, United Arab Emirates.}%
}

\markboth{}%
{Shell \MakeLowercase{\textit{et al.}}: Bare Demo of IEEEtran.cls for IEEE Journals}
%



\maketitle

\begin{abstract}
Thermal management is a critical challenge in compact computing platforms operating under sustained high computational loads. Elevated processor temperatures can lead to thermal throttling, degraded performance, and reduced reliability. This study develops and experimentally evaluates a smart hybrid heat sink that integrates passive thermal energy storage with active cooling for sustained device operation. The proposed design comprises an aluminium cavity containing paraffin wax as a phase change material (PCM) embedded within a copper foam and integrated with an adaptively controlled fan. {\color{black}Fan operation is guided by a Random Forest regression model trained on experimentally measured CPU temperature data to predict future CPU temperatures. The predicted temperature is then used to support the adaptive fan-control strategy.} The heat sink is mounted directly on the CPU of a Raspberry Pi 5 and experimentally evaluated under an extreme CPU workload of approximately 99\% CPU utilization. The performance of the proposed hybrid system is compared against two benchmark scenarios: passive cooling using PCM embedded in copper foam and a commercial Raspberry Pi 5 active cooler. The proposed heat sink demonstrates superior thermal performance to both benchmarks, achieving peak CPU temperature reductions of approximately \SI{9}{\celsius} relative to the PCM–copper foam configuration and \SI{6}{\celsius} relative to the commercial active cooler.

\end{abstract}


\IEEEpeerreviewmaketitle

\section{Introduction}

The rapid growth of digital technologies and artificial intelligence (AI) has increased the computational demands placed on compact electronic devices. Higher processing capability within increasingly confined spaces leads to greater power densities and localized heat generation, making effective thermal management essential for maintaining performance and reliability. These challenges are particularly important for compact computing platforms such as the Raspberry Pi 5, where sustained computational loads can drive the processor toward its thermal-throttling limits in the absence of adequate cooling \cite{allan2023raspberrypi}.

Conventional thermal-management approaches generally rely on passive or active cooling. Among passive techniques, phase change materials (PCMs) have received considerable attention because of their ability to absorb heat as latent energy during melting, thereby delaying temperature rise and reducing transient temperature fluctuations. Mallela et al. \cite{mallela2025pcm} experimentally demonstrated the thermal benefit of this approach, reporting nearly 6\% lower peak temperatures using PCM-filled aluminum fin heat sinks. Kumar et al. \cite{kumar2022paraffin} experimentally examined paraffin-based PCM heat sinks with cross-plate fins and reported reductions in heat-sink temperature of up to 19.5\% by increasing the number of cavities within the PCM heat sink. 

Despite these advantages, PCM cooling is constrained by its finite thermal-storage capacity and, for commonly used paraffin-based PCMs, relatively low thermal conductivity. Once the PCM becomes fully melted, its ability to absorb additional heat is substantially reduced, while low thermal conductivity can lead to nonuniform melting. Metallic foams have therefore been investigated to enhance the thermal response of PCM-based heat sinks. Afaynou et al. \cite{afaynou2025hybrid} reported that fully filling a PCM heat sink with aluminum foam reduced the electronic-component temperature by 16.3 °C and the PCM melting time by 13.85\% compared with an unfilled heat sink. Arshad et al. \cite{arshad2022pcmfoam} investigated PCM–metal-foam heat sinks for electronic cooling and showed that the incorporation of copper foam improved heat-transfer performance, with an optimized configuration reducing PCM melting time by 18.1\%. Although these enhancements improve the thermal response of PCM-based heat sinks, the finite thermal-storage capacity of the PCM remains a limitation during prolonged operation.

In parallel, machine-learning techniques have increasingly been applied to thermal-management systems for prediction and control. 
Gürsoy et al. \cite{gursoy2025latent} applied an LSTM-Attention model to PCM–metal-foam thermal storage systems and achieved ($R^2>0.997$) and a mean absolute percentage error below 3\% in predicting PCM thermal characteristics. Similarly, Sreenath and Chanda \cite{sreenath2025modelling} developed an artificial-neural-network model using experimental data from a PCM–metal-foam composite heat sink and achieved ($R^2>0.99$) in predicting the heat-sink interface temperature under different heat inputs. In such studies, machine learning was primarily used to predict or optimize thermal-system performance. Predictive control has also been investigated in actively cooled electronic systems. Iranfar et al. \cite{iranfar2020dynamic} developed a reinforcement-learning-based thermal-management strategy that proactively adjusted processor operating conditions and fan speed, reducing fan power by up to 40\% relative to constant fan-speed operation. Chen et al. \cite{chen2023adaptive} used machine learning to predict future processor temperature and reinforcement learning to determine the subsequent thermal-control action, improving system performance by up to 43\%. However, limited attention has been given to integrating predictive active cooling with PCM–metal-foam thermal buffering, where the predicted future processor temperature directly governs fan operation. To this end, this study develops and experimentally evaluates a compact hybrid cooling system for the Raspberry Pi 5 that integrates paraffin PCM, copper foam, and prediction-driven forced convection.

\section{Problem Setup}

We employ a proactive compact hybrid cooling assembly designed for the Raspberry Pi 5, intended to mitigate or avoid thermal throttling during sustained, computationally intensive workloads. Heat generated by the system-on-chip (SoC) is transferred to an aluminum heat sink containing paraffin PCM embedded within copper foam. The PCM absorbs transient thermal loads through latent heat storage, while the copper foam enhances heat transfer within the PCM. A variable-speed fan provides forced convection to reject heat from the assembly and support sustained operation once passive thermal buffering becomes insufficient.

\begin{figure}[htbp]
    \centering
    \includegraphics[width=0.5\linewidth]{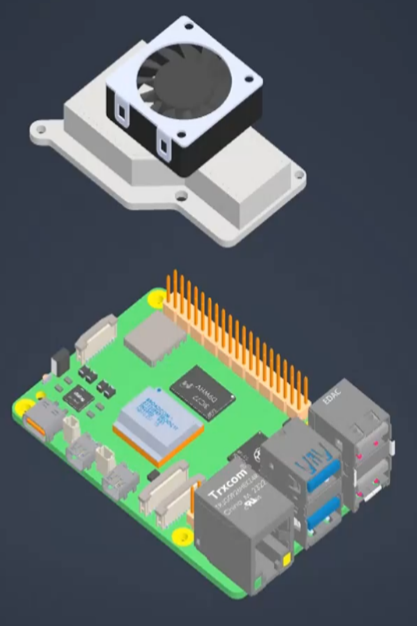}
    \caption{CAD model of the proposed hybrid cooling system integrated with the Raspberry Pi 5.}
    \label{fig:CAD}
\end{figure}

The cooling assembly is mounted directly above the Raspberry Pi 5 processor, as illustrated in Fig.\ref{fig:CAD}. The system is designed such that fan operation can be adjusted based on the predicted thermal response of the processor, allowing passive thermal buffering and active cooling to operate as an integrated thermal-management system.

\section{System Design}

\subsection{Mechanical Subsystem}

The cooling assembly consists of a CNC-machined aluminum heat sink mounted directly on the Raspberry Pi 5 processor. A thermal interface pad is placed between the processor and the aluminum base to reduce thermal contact resistance, allowing heat generated by the SoC to be conducted into the heat-sink body and the internal PCM cavity.

The internal cavity contains paraffin wax PCM with a melting range of 53--57$^\circ$C and latent heat capacity of 200 kJ/kg. The melting range was selected so that the PCM begins absorbing latent heat before the processor approaches its thermal-throttling range. A copper foam insert with a porosity of 95\% and 50 pores per inch (PPI) is embedded within the PCM. The high porosity allows the PCM to occupy most of the cavity volume while the copper structure improves heat spreading within the PCM region.

\begin{figure}[htbp]
    \centering
    \includegraphics[width=0.25\textwidth]{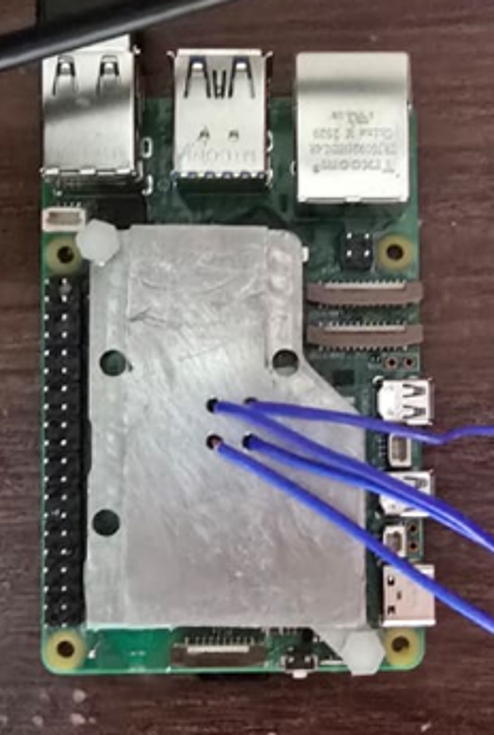}
    \caption{Aluminum PCM cavity and lid with thermocouple openings.}
    \label{fig:Thermocouple}
\end{figure}

The PCM cavity is enclosed by an aluminum lid that incorporates openings for temperature measurement (confer Fig.\ref{fig:Thermocouple}). During assembly, the copper foam is first positioned inside the cavity and the PCM is then introduced to fill the void space within the foam before the lid is installed.

A PWM-controlled fan is mounted above the heat-sink assembly to provide forced-convection cooling during sustained operation. The fan is connected to the Raspberry Pi and its operating speed is adjusted by the predictive control system described in the following sections.

Temperature measurements within the cooling assembly are obtained using K-type thermocouples with an accuracy of $\pm 0.9^\circ$C. The thermocouples are inserted through fixed openings in the lid and positioned at approximately 15\%, 50\%, and 80\% of the PCM cavity height, enabling the temperature development through the PCM region to be monitored during operation (see Fig.\ref{fig:Thermocouple}).


\subsection{System Architecture}
We adopt a modular architecture organized around four responsibilities: sensing, prediction, control, and data management. All modules run on a single controller node, the Raspberry Pi 5, {\color{black}which acquires temperature measurements from the cooling-system sensors}, determines an appropriate fan speed, drives the fan through the standard header, and records a time-stamped run log for subsequent analysis. This single-node architecture simplifies integration,  minimizes communication latency, and maintains modularity so that individual components can be updated without affecting the rest of the system.

At the node level, the system comprises a controller node interfaced with two physical peripherals: temperature sensors positioned on and around the cooling assembly, and a variable-speed fan mounted above the sealed PCM cavity. All computations are performed locally on the controller, including sensor data acquisition, fan-speed selection, and run-log storage. This configuration allows straightforward deployment on any Raspberry Pi 5 based platforms.


At the component level, the controller node comprises several cooperating software modules.
The Sensing Module acquires temperature measurements and basic operating data at a fixed sampling interval, followed by lightweight validation. The Feature/Window Module maintains a short history of recent measurements as input to the predictor. {\color{black} The ML Predictor uses this history to predict the future CPU temperature. The Thermal Controller uses the predicted temperature, together with predefined temperature bands and safety limits, to determine the appropriate fan setting.} 
The Fan Driver applies the selected setting through the standard fan header. Finally, the Data Store/Logger appends time-stamped records of measurements, fan settings, and any flagged events to the run log for subsequent analysis and comparison across tests.
\fh{Data acquisition and preprocessing
•	Model development 
•	Predictive fan control strategy }

\subsection{Controller Architecture}
\label{Sec:controller}

Following the thermal-management design, the Raspberry Pi 5 acts as both the device under test and the cooling controller. 
The control software performs four functions: temperature acquisition, fan-setting selection, fan actuation, and event logging for subsequent model refinement. The controller operates in a continuous loop. First, it samples the temperatures representing the SoC and the cooling assembly followed by basic validity checks. Second, it chooses an appropriate fan speed using two mechanisms: (i) a simple rule-based scheme in which discrete temperature bands are combined with a hysteresis margin to prevent frequent switching between fan-speed levels and {\color{black}(ii) a machine-learning-based prediction of the future CPU temperature based on the recent thermal profile.} 

The rule-based controller selects the fan speed from predefined temperature bands. Hysteresis is applied at the band boundaries to prevent rapid switching between speed levels. Explicit safety limits are enforced regardless of the predictive output, so the SoC temperature stays below the \SI{85}{\celsius} steady-state limit even if the model misbehaves or a sensor fails. 

{\color{black}A Random Forest regression model is employed to predict the future CPU temperature. It uses a short window of recent measurements as input and predicts the CPU temperature 30 s ahead. The predicted temperature is then supplied to the thermal controller, which determines the appropriate fan setting in conjunction with the rule-based logic and predefined safety limits.}
It was trained on experimentally measured temperature data from the Raspberry Pi 5 under varying workloads. This model was chosen for its ability to capture nonlinear thermal dynamics and its robustness to measurement noise. In addition, it is lightweight, which makes it well suited for deployment on the Raspberry Pi 5. Multiple model configurations were evaluated using metrics such as time-to-throttle, maximum SoC temperature, average fan setting, noise and data completeness and the configuration with the best predictive performance was selected to drive the fan.

{\color{black}Finally, the control decision is subject to predefined safety limits. If the measured temperature exceeds a critical threshold or an invalid sensor reading is detected, the controller overrides the predictive output, switches the fan to maximum cooling, and logs the event for subsequent analysis.}

\section{Results and Discussion}
\fh{•	Machine learning prediction and performance (the below points are suggestions, please add where you find necessary).
--Temperature prediction accuracy 
--Adaptive fan response 
}

The performance of the proposed hybrid cooling system was evaluated in terms of thermal response and temperature-prediction accuracy under an extreme workload of approximately 99\% CPU utilization. All experiments were repeated three times, and the averaged responses are presented. The thermal performance of the proposed system was compared with passive PCM–metal-foam cooling and the commercial Raspberry Pi 5 active cooler.

\subsection{Thermal Throttling Under Extreme Workload}

\begin{figure}[htbp]
    \centering
    \includegraphics[width=\columnwidth]{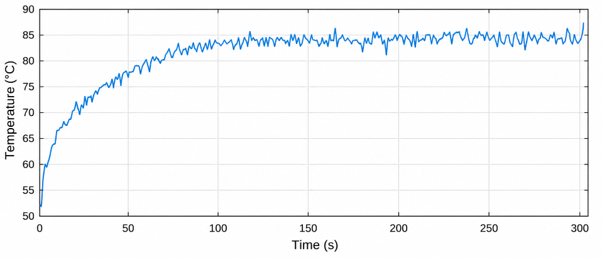}
    \caption{CPU temperature under the extreme workload without cooling.}
    \label{fig:no_cooling_temperature}
\end{figure}

\begin{figure}[htbp]
    \centering
    \includegraphics[width=\columnwidth]{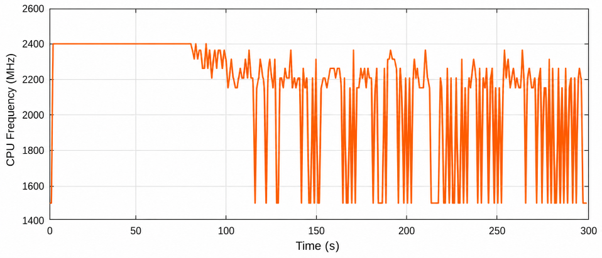}
    \caption{CPU frequency under the extreme workload without cooling.}
    \label{fig:no_cooling_freq}
\end{figure}

Fig.\ref{fig:no_cooling_temperature} illustrates the CPU temperature response of the Raspberry Pi 5 under the extreme workload without additional cooling. The CPU temperature increased rapidly and approached the thermal-throttling range, stabilizing at approximately 85°C. Fig.\ref{fig:no_cooling_freq} illustrates the corresponding CPU-frequency response. The frequency remained at approximately 2400 MHz during the initial stage of the test before exhibiting repeated reductions after approximately 75 s, indicating the onset of thermal throttling.

\subsection{Machine Learning Performance}
\begin{figure}[htbp]
    \centering
    \includegraphics[width=\columnwidth]{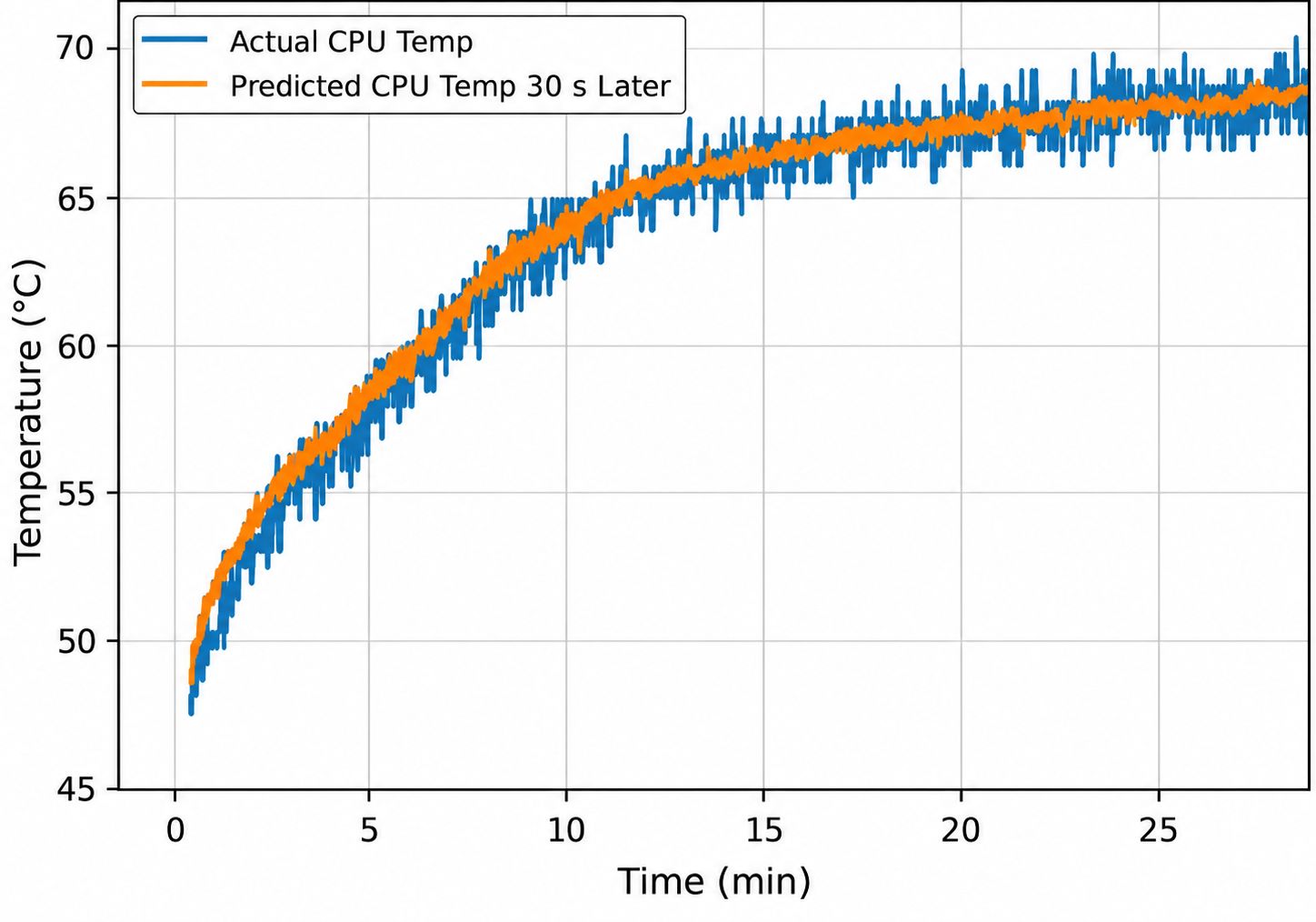}
    \caption{Machine learning model prediction performance}
    \label{fig:ML_predictions}
\end{figure}

Fig.\ref{fig:ML_predictions} illustrates the measured CPU temperature and the corresponding 30-s-ahead temperature predicted by the Random Forest model. The predicted response closely follows the measured thermal trend over the duration of the experiment. The selected model achieved an \(R^2\) value of 0.931, a mean absolute error (MAE) of 1.076°C, and a root mean square error (RMSE) of 1.946°C. {\color{black}These results indicate that the model captures the processor temperature evolution with good agreement between the measured and predicted responses, supporting its use within the predictive fan-control strategy.}

\subsection{Thermal Performance of the Hybrid Cooling System}

\begin{figure}[htbp]
    \centering
    \includegraphics[width=\columnwidth]{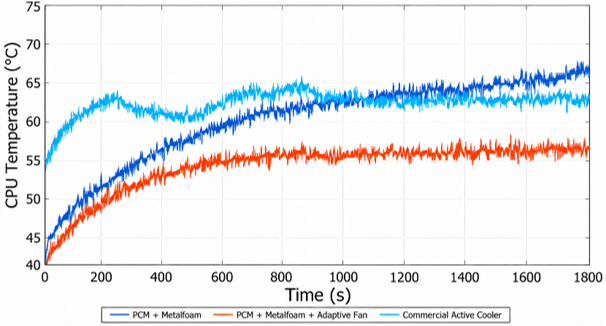}
    \caption{CPU temperature under the extreme workload with cooling.}
    \label{fig:CPU_temperature}
\end{figure}

Fig.\ref{fig:CPU_temperature} compares the CPU temperature responses of the proposed hybrid system, the passive PCM–metal-foam configuration, and the commercial Raspberry Pi 5 active cooler under the same extreme workload. The proposed hybrid system maintained an average CPU temperature of approximately 57°C and a peak temperature of 59.5°C. In comparison, the commercial active cooler reached a peak temperature of approximately 65°C, while the passive PCM–metal-foam configuration exhibited a higher peak temperature. The proposed system therefore reduced the peak CPU temperature by approximately 6°C relative to the commercial active cooler and by approximately 9°C relative to the passive PCM–metal-foam configuration. No thermal throttling was observed with the proposed system, and the CPU frequency remained stable throughout the test, indicating sustained processor operation under the extreme workload.

\section{Conclusion}
This study developed and experimentally evaluated a compact hybrid cooling system for the Raspberry Pi 5 that integrates paraffin PCM, copper foam, and prediction-driven forced convection. Under an extreme workload of approximately 99\% CPU utilization, the Raspberry Pi 5 without additional cooling reached the thermal-throttling region after approximately 75 s, with the CPU temperature stabilizing near 85°C. The Random Forest model predicted the CPU temperature 30 s ahead with an \(R^2\) of 0.931, a mean absolute error of 1.076°C, and a root mean square error of 1.946°C. When integrated with the adaptive fan-control strategy, the proposed hybrid system maintained an average CPU temperature of approximately 57°C and a peak temperature of 59.5°C without thermal throttling. This corresponds to peak-temperature reductions of approximately 6°C relative to the commercial Raspberry Pi 5 active cooler and 9°C relative to the passive PCM–metal-foam configuration. These results demonstrate that combining passive thermal buffering with prediction-driven active cooling can provide stable thermal management for sustained high-load operation in compact computing platforms.

\section*{Acknowledgment}
Research reported in this manuscript was supported by an
undergraduate research grant from the American University of Sharjah (grant number: UE2619);


\ifCLASSOPTIONcaptionsoff
  \newpage
\fi



%
\bibliographystyle{IEEEtran}
\bibliography{references}

%
%

%







\end{document}